# Does your surname affect the citability of your publications?[1]

Giovanni Abramo (*corresponding author*)
  *Laboratory for Studies in Research Evaluation*
  *Institute for System Analysis and Computer Science (IASI-CNR)*
  *National Research Council of Italy*
   ADDRESS:   Istituto di Analisi dei Sistemi e Informatica, Consiglio Nazionale delle Ricerche
              Via dei Taurini 19, 00185 Roma - ITALY
              tel. +39 06 7716417, fax +39 06 7716461,
              giovanni.abramo@uniroma2.it

Ciriaco Andrea D'Angelo
  *University of Rome "Tor Vergata" and Institute for System Analysis and Computer Science-National Research Council of Italy*
   ADDRESS:   Dipartimento di Ingegneria dell'Impresa, Università degli Studi di Roma "Tor Vergata"
              Via del Politecnico 1, 00133 Roma - ITALY
              Tel. and fax +39 06 72597362, dangelo@dii.uniroma2.it

**Abstract**

Prior investigations have offered contrasting results on a troubling question: whether the alphabetical ordering of bylines confers citation advantages on those authors whose surnames put them first in the list. The previous studies analyzed the surname effect at publication level, i.e. whether papers with the first author early in the alphabet trigger more citations than papers with a first author late in the alphabet. We adopt instead a different approach, by analyzing the surname effect on citability at the individual level, i.e. whether authors with alphabetically earlier surnames result as being more cited. Examining the question at both the overall and discipline levels, the analysis finds no evidence whatsoever that alphabetically earlier surnames gain advantage. The same lack of evidence occurs for the subpopulation of scientists with very high publication rates, where alphabetical advantage might gain more ground. The field of observation consists of 14,467 scientists in the sciences.

---

# 1. Introduction

Questions concerning authors' surnames have attracted attention in the scientometric literature for some time. Among other issues, the surname initial is considered as a possible influencing factor on citability and the individual's academic career. Advantages from a name coming earlier in the alphabet could be expected in the disciplines with the alphabetical ordering tradition for the author byline.

In fact there is a substantial degree of heterogeneity in the manner in which author orderings are assigned, both across and within disciplines (Joseph, Laband, & Patil, 2005). Empirical analysis revealed that the use of alphabetical ordering in scientific publishing is declining over time (Waltman, 2012). The use of such ordering seems most common in mathematics, economics (including finance), and high energy physics. Instances of "alphabetical discrimination" could then arise because of several reasons. First of all, the common use of the "et al." citation rule in the body of the text could immediately establish a higher level of attention to the first author. In this regard, there is also the fact that certain citation indices and research engines, such as Econlit in the field of economics, follow a long-established practice of registering only the first surnames in the byline (van Praag & van Praag, 2008). Second, in the case of evaluators originating from disciplines with contribution-based name ordering, the evaluator could then associate a higher contribution to the individuals listed first in the byline solely on the basis of their name (Levitt & Thelwall, 2013; van Praag & van Praag, 2008). Third, as the reference lists at the close of the articles are usually ordered alphabetically, it is likely that the first-placed authors will receive greater recognition and be easier to remember.

The literature seems to confirm that scientists are generally aware of the potential of discrimination due to alphabetized bylines. Kadel and Walter (2015) examined the co-authorship behavior in economics and showed that scholars late in the alphabet refrain from publishing articles with three or more authors. However, no such evidence was found in the field of finance. Some researchers even seem to manipulate the use of their surnames to obtain a higher position in the byline, for example by choosing between more than one last name, by using the family name with or without prefix, or through the transcription of Greek names into English (Efthyvoulou, 2008).

Several studies have analyzed whether the surname position in the alphabet can affect an individual's academic career. Einav and Yariv (2006) found that the probabilities of receiving tenure at top-5 and top-10 American economics departments are roughly one percentage point higher per letter earlier in the alphabet. At the same time, their analysis revealed no significant correlation between the first letter of surnames and tenure status in psychology, where authors follow the convention of organizing the byline according to the authors' contributions. Efthyvoulou (2008) revealed that the probability of being employed in a highly considered U.S. economics department is 21.5% higher for "A-professors" than it is for "Z-professors". The same author also considered the number of downloads and page views on research networks as a measure of success. Regarding this, he detected that for individuals with last name initial "A", compared to "Z", there was a nearly 58% increase in the probability of being among those authors with the most downloaded papers. Similarly, having a name beginning with A rather than Z resulted in a 39% increase in the probability of being among the authors whose works received the highest attention.

Van Praag and van Praag (2008) examined individuals' surname initials relative to publication rates in 11 mainstream economics journals for a sample of highly productive economists, finding that there was a significant relationship between the two. Their estimates indicate that a Z-author should be considered as deserving a 16% "premium" in the compensation for his or her performance, measured in average yearly publications, if compared directly to an A-author. Arsenault and Larivière (2015) showed that "uncitedness rates tend to increase with the progression of the first author's last name in the alphabet indicating that papers with a first author whose last name starts with a letter that occurs later in the alphabet might be less visible". They further observed that the phenomenon is more noticeable in Mathematics and Physics, where the alphabetical ordering practice for the author byline is well established.

To the best of our knowledge, only three studies have focused on the specific question of the effect of the surname initial on publications' citability. Shevlin and Davies (1997) were the first to investigate the phenomenon. They analyzed the 1994 Web of Science (WoS) Science Citation Index, and found no effect. However the citations in this study were not field-normalized and we also note that the citation window was too short for a reliable prediction of the definitive impact of publications. Huang (2015) analyzed a WoS 1990-2005 extract of 846,122 U.S.-authored papers in 12 fields of the sciences. The results were that papers where the first-listed author has an alphabetically "early" surname are cited more frequently, but that there is no distinction in citations when considering the names of the other authors of the publication. In particular, "the estimates show that shifting surname initial from the bottom of the alphabet to the top is associated with a 0.44 percentile increase in rank of citations among the papers published in the same year," indicating a sizable alphabetical bias. Ong et al. (2015) applied the same methodology as Huang to the fields of economics and management, and found "a positive citation rank trend on author initials for single-authored economics papers when compared with management single-authored papers, alone and as compared to economics coauthored papers".

These three studies analyze the surname effect at publication level, i.e. whether papers with the first author early in the alphabet trigger more citations than papers with a first author late in the alphabet. We adopt instead a different approach, by analyzing the surname effect on citability at the individual level, i.e. whether authors with alphabetically earlier surnames result as being more cited. The rationale is that if alphabetical order leads to an increased visibility of authors with earlier surnames, this should affect all their output and thus their overall citability. The research question is not at all trivial, given that it touches on the careers of hundreds of thousands of academics. Should the surname effect reveal true and significant, then the matter should be taken into consideration in the assessment of research activities, at both individual and group levels, in the disciplines still applying simple alphabetical order for publication bylines.

We consider all of the publications by all professors in sciences of Italian universities, with the exception of the life sciences where standard practice in Italy is to order the byline according to the authors' intellectual contributions. We examine the link between the first letter of the surname and the field-normalized citations received, at the overall level and the discipline level. We then check for differences in these links when the dataset is restricted to "top" scientists, defined as those with the highest number of publications. The reason for this second analysis is that the higher the number of publications by the scientists, the more evident the surname effect should be.

We restrict our analysis to Italy only for three reasons: i) we can rely on highly accurate disambiguated data at individual level; ii) in Italy, alphabetical authorship is a relatively common phenomenon in many fields; (3) we do not have access to WoS data for other countries. In fact, we have disambiguated all the authorships of WoS indexed publications by professors of Italian universities since 2001. Knowing the entire publication portfolio of each professor, we are able to assess the effect of the surname initial on citability of their overall scientific output, not just of that falling in the extracted publication sample or in the particular discipline under study[2]. The restriction to Italy, while making the analysis much more precise, does not jeopardize the generalization of results, as Italian articles are cited from all over the world and not just by Italian scholars. All others equal, there is no reason why foreign scholars should cite more Italian authors with surname initials earlier in the alphabet, while being indifferent to the surname initial of authors from other countries.

Furthermore, we conduct the analysis at the individual level, differently from previous studies that, without author's name disambiguation, need to group publications by surname initial. The difference is subtle, but not negligible. We show it with an example. Let us assume that 20 out of the 24 publications by the three authors of the population whose surname initial is A do show citation gains, vis-à-vis other publications by authors with non-A initials. The conclusion would be that surname initials affect citability. Through author's name disambiguation though, we are able to find out that the 20 publications with citation gains all belong to one author, and the remaining four to the other two authors. The analysis at the individual level then would lead to the opposite conclusion. Differently from previous studies, what we investigate is the effect of surname initial on the citability of (all the publications of) individual authors[3], not on the publications (only co-authored ones) aggregated by surname initial. It is the former in fact that may affect authors' behavior to cope with a supposed "alphabetical discrimination".

In the next section of the paper we describe the dataset and methods. The third section presents the results of the analysis, and the final section the conclusions.

## 2. Data and methods

The field of observation consists of all professors of Italian universities in the science disciplines, except those in the life sciences, where standard practice in Italy is to order the byline according to the authors' intellectual contributions. We observe the professors' research outputs as indexed in WoS in the period 2001-2004. For reasons of significance we consider only those professors with at least three years of employment over the period 2001-2004.

We count the citations at the end of 2015, thus allowing a citation window wide enough for any discriminating effect of alphabetical ordering to appear. The substantial window also makes the citation counts more reliable in serving as a predictor of the real impact of publications.

---
[2] It is not infrequent that scientists publish also outside their field of specialization. For example, just think of a statistician, whose publication output can spread across such different disciplines as economics, medicine, agricultural sciences, etc.; or of multidisciplinary teams whose research output does not necessarily fall in all the fields that the team members represent.

[3] If higher attention because of surname initial might be conducive of higher citability, then one should expect that higher citability occurs in general for all publications of an author, not just for those where they are first authors.

In the Italian university system, each professor is classified in one and only one research field. In the sciences, there are 205 such fields (named "scientific disciplinary sectors", SDSs[4]), grouped into nine disciplines (named "university disciplinary areas", UDAs). The nine UDAs are: mathematics and computer sciences; physics; chemistry; earth sciences; biology; medicine; agricultural and veterinary sciences; civil engineering; industrial and information engineering. Table 1 reports in each UDA the percentages of publications whose first author is alphabetically ordered. The last column shows the expected value considering a completely random distribution of the authors in the byline.[5] The practice of ordering authors by their contribution is evident in the life sciences (above all, UDA 5 and 6), while the opposite reveals true in mathematics and computer sciences. In the remaining disciplines mixed practices occur, even if the observed percentage of publications with first author alphabetically ordered is significantly higher than the expected one considering a systematic random ordering of the byline. While earlier studies often focused on social sciences fields, in particular economics and finance, we have nevertheless chosen to exclude them, because it turns out that Italian professors in these fields have a too large share of their publication output in sources that are not covered by Web of Science.[6] Findings then would not be robust enough. It must be noted that the faculty staff in the sciences represents above 60% of total staff, while social scientists only 20%.

*Table 1: Percentages of co-authored publications by Italian universities whose first author is alphabetically ordered. 2001-2004 WoS publications.*

| UDA* | No. of co-authored publications | Average no. of co-authors | Percentage of publications with first author alphabetically ordered | |
|---|---|---|---|---|
| | | | Observed | Expected value |
| 1 | 6,111 | 2.74 | 89.2% | 40.1% |
| 2 | 11,300 | 4.54 | 46.0% | 26.9% |
| 3 | 12,574 | 4.83 | 43.7% | 24.1% |
| 4 | 2,214 | 4.03 | 42.7% | 29.5% |
| 5 | 10,803 | 5.07 | 24.6% | 23.7% |
| 6 | 21,996 | 6.13 | 19.9% | 19.3% |
| 7 | 3,974 | 4.52 | 29.4% | 25.8% |
| 8 | 1,998 | 2.87 | 63.1% | 38.4% |
| 9 | 15,098 | 3.47 | 61.5% | 32.9% |

\* 1=Mathematics and computer science; 2=Physics; 3=Chemistry; 4=Earth sciences; 5=Biology; 6=Medicine; 7=Agricultural and veterinary sciences; 8=Civil engineering; 9=Industrial and information engineering

---

[4] The complete list is accessible on http://attiministeriali.miur.it/UserFiles/115.htm, last accessed 05/12/2016.

[5] To calculate the expected value, first we randomly order the co-authors of each publication, and then tally the publications whose first author's surname begins with a letter earlier than the other co-authors' in the byline.

[6] The percentages of Italian professors (by field) who have none of their outputs in the period under observation covered by WoS, are: Political economy, 66.2%; Economic policy, 75.0%; Finance, 69.2%; History of economic thought, 86.7%; Econometrics, 28.0%; Applied economics, 77.4%; Business administration, 96.0%; Corporate finance, 87.2%; Financial management, 100.0%; Business organisation, 81.4%; Economics of financial intermediaries, 95.3%; Economic history, 95.3%; Commodity studies, 67.9%.

The classical Italian alphabet is composed of only 21 letters, with the letters J, K, W, X, and Y being foreign letters of only recent addition, thus hardly appearing in family names. Furthermore, the letter *h* is silent, therefore Italian surnames beginning with "H" are virtually unknown. Names beginning with these letters either indicate Italian professors with foreign origins on the father's side, or more likely individuals of foreign nationality. Excluding such very few surnames (66 in all) from the analysis, we have a dataset composed of 14,467 professors. The source for data on the faculty at each university is the database maintained by the Ministry of Education, Universities and Research,[7] which indexes the names, gender, academic rank, field (SDS/UDA), and institutional affiliation of all professors in Italian universities at the end of each year.

The distribution of the dataset per UDA and first letter of surname is shown in Table 2. We notice that the top five letters represent 34.3% of the population, while the bottom five, 9.7%.

The data on the professors' publications were extracted from the Observatory of Public Research in Italy (ORP),[8] a database derived under license by the authors from the Thomson Reuters WoS National Citation Report for Italy. Taking the raw WoS data, and applying complex algorithms for the reconciliation of the author's affiliation and the disambiguation of their true identity, each publication is attributed to the professors who authored it (D'Angelo, Giuffrida, & Abramo, 2011). The harmonic average of precision and recall (F-measure) of authorships disambiguated by our algorithm is around 97% (2% sampling error, 98% confidence interval).

We approximate the impact of the publications by citation counts. We also normalize the citations by scientific field. This avoids the distortions otherwise caused by variations in citation behavior across fields (Abramo, Cicero, & D'Angelo, 2012a). We use the indicator called "Article Impact Index" (AII), calculated as the ratio of the number of citations received by the publication, to the average of the citations for all cited Italian publications[9] of the same year and WoS journal subject category.

Once each professor's publication portfolio for the observation period is identified we then calculate the average AII per paper of the portfolio ($\overline{AII}$). This indicator is similar to the famous "Mean Normalized Citation Score" (MNCS) (Waltman et al., 2011), introduced by the Leiden group, except for the minor differences in the operationalization of the measurement, as described in Abramo & D'Angelo (2016). Finally, we examine whether a link exists between the value of $\overline{AII}$ recorded for the different professors and their surname initials.

The next section presents the results of the analyses.

---

[7] http://cercauniversita.cineca.it/php5/docenti/cerca.php, last accessed 05/12/2016.

[8] www.orp.researchvalue.it, last accessed 05/12/2016.

[9] Abramo, Cicero, and D'Angelo (2012b) demonstrated that the average of the distribution of citations received for all cited publications of the same year and WoS journal subject category is the most effective scaling factor.

*Table 2: Distribution of professors per surname initial in each UDA*

| Surname initial / UDA* | 1 | 2 | 3 | 4 | 7 | 8 | 9 | Total |
|---|---|---|---|---|---|---|---|---|
| A | 79 | 79 | 107 | 17 | 73 | 39 | 140 | 534 |
| B | 234 | 244 | 270 | 91 | 204 | 98 | 340 | 1481 |
| C | 243 | 248 | 398 | 119 | 247 | 130 | 446 | 1831 |
| D | 159 | 149 | 205 | 62 | 133 | 80 | 252 | 1040 |
| E | 11 | 16 | 16 | 7 | 12 | 2 | 12 | 76 |
| F | 134 | 118 | 167 | 46 | 108 | 60 | 189 | 822 |
| G | 160 | 148 | 221 | 55 | 148 | 64 | 233 | 1029 |
| I | 16 | 19 | 29 | 11 | 23 | 10 | 46 | 154 |
| L | 108 | 88 | 120 | 44 | 96 | 47 | 174 | 677 |
| M | 251 | 254 | 334 | 103 | 226 | 119 | 384 | 1671 |
| N | 36 | 39 | 42 | 13 | 38 | 24 | 65 | 257 |
| O | 24 | 25 | 27 | 9 | 16 | 11 | 31 | 143 |
| P | 197 | 222 | 279 | 81 | 190 | 83 | 318 | 1370 |
| Q | 2 | 4 | 8 | 2 | 8 | 1 | 12 | 37 |
| R | 112 | 125 | 152 | 61 | 115 | 74 | 186 | 825 |
| S | 164 | 187 | 210 | 82 | 151 | 71 | 260 | 1125 |
| T | 93 | 85 | 123 | 45 | 73 | 33 | 135 | 587 |
| U | 7 | 3 | 14 |  | 5 | 5 | 11 | 45 |
| V | 77 | 76 | 88 | 31 | 69 | 45 | 107 | 493 |
| Z | 42 | 40 | 58 | 18 | 44 | 17 | 51 | 270 |
| Total | 2,149 | 2,169 | 2,868 | 897 | 1,979 | 1,013 | 3,392 | 14,467 |

* *1=Mathematics and computer science; 2=Physics; 3=Chemistry; 4=Earth sciences; 7=Agricultural and veterinary sciences; 8=Civil engineering; 9=Industrial and information engineering*

## 3. Results

Bibliometric indicators present typical power law distributions. In effect, the Shapiro-Wilk test rejects the hypothesis of normality of $\overline{AII}$ distributions. Given this, to better describe the data concerning the indicator, Table 3 presents the reference quartiles for each surname initial, while Figure 1 provides a graph of the median values and interquartile ranges.

The median values of $\overline{AII}$ vary from a minimum of 0.486 per professors with initial letter N and a maximum of 0.682 for those with initial Q, both of which are letters with low numbers of observations. On the other hand, the first letters most frequently present show median values that differ very little: the seven letters with more than 1000 observations each (B, C, D, G, M, P and S) show a median $\overline{AII}$ varying from a minimum of 0.534 to a maximum of 0.579. The dispersion of values also appears quite similar for all 20 groups: the interquartile range varies between 0.6 and 0.7 for the distributions referring to 17 of the surname initials, and arrives at a maximum of 0.805 for the distribution referring to the initial O. The last column of Table 3 invariably shows a minimum value of nil $\overline{AII}$, indicating that there is at least one professor in each group who has not received any citations. On the other hand, the maximum value of $\overline{AII}$ fluctuates between 3.0, for the surname initials E, Q and U, and 28.7 for first letter D.

*Table 3: Descriptive statistics of $\overline{AII}$ per surname initial of professors in the dataset*

| Letter | Obs | I quartile | Median | III quartile | Interquartile range | Range (min-max) |
|--------|-----|-----------|--------|--------------|---------------------|-----------------|
| A | 534 | 0.275 | 0.555 | 0.920 | 0.645 | [0-4.022] |
| B | 1481 | 0.272 | 0.547 | 0.904 | 0.632 | [0-5.977] |
| C | 1831 | 0.240 | 0.549 | 0.925 | 0.685 | [0-16.696] |
| D | 1040 | 0.253 | 0.534 | 0.879 | 0.626 | [0-28.720] |
| E | 76 | 0.253 | 0.521 | 0.880 | 0.628 | [0-2.481] |
| F | 822 | 0.239 | 0.524 | 0.881 | 0.642 | [0-6.265] |
| G | 1029 | 0.257 | 0.553 | 0.905 | 0.649 | [0-10.708] |
| I | 154 | 0.278 | 0.528 | 0.883 | 0.605 | [0-4.789] |
| L | 677 | 0.240 | 0.527 | 0.931 | 0.691 | [0-13.341] |
| M | 1671 | 0.275 | 0.579 | 0.959 | 0.683 | [0-7.377] |
| N | 257 | 0.213 | 0.486 | 0.869 | 0.656 | [0-3.368] |
| O | 143 | 0.212 | 0.564 | 1.017 | 0.805 | [0-19.220] |
| P | 1370 | 0.264 | 0.546 | 0.913 | 0.650 | [0-7.184] |
| Q | 37 | 0.338 | 0.682 | 0.980 | 0.642 | [0.025-2.738] |
| R | 825 | 0.286 | 0.569 | 0.932 | 0.646 | [0-7.273] |
| S | 1125 | 0.258 | 0.537 | 0.977 | 0.720 | [0-12.163] |
| T | 587 | 0.230 | 0.544 | 0.908 | 0.678 | [0-8.262] |
| U | 45 | 0.195 | 0.623 | 0.947 | 0.753 | [0-2.615] |
| V | 493 | 0.278 | 0.571 | 0.950 | 0.672 | [0-5.587] |
| Z | 270 | 0.282 | 0.592 | 0.932 | 0.649 | [0-5.786] |

This first summary analysis shows distributions of average standardized citations that seem not particularly dissimilar per surname initial. However observing Figure 1, we are left with the impression that the initials in the second half of the alphabet, while showing more variability of distribution, also present slightly higher median values of $\overline{AII}$. If true, a regression network interpolating these twenty points would have positive slope. In effect, the Spearman correlation coefficient between alphabetic rank[10] and the median value of the relative distributions of $\overline{AII}$ is equal to -0.423, which would suggest backing for the surprising hypothesis of a "disadvantage" for those surnames beginning with alphabetically early letters: a completely new understanding with respect to the existing literature. In reality this is a hasty and erroneous conclusion, because it is above all the letters with a low number of observations (Q,U,Z) that lead to this type of inferential result. Controlling for the frequency of the first letters of name, i.e. correlating the two variables (first letter rank vs $\overline{AII}$ rank) for all 14,467 observations, the Spearman ρ drops to -0.012, leading us to reject any sort of hypothesis of an across-the-board association between the citability of a scientist's works and the rank of their surname initial.

---

[10] We assigned to surname initials A the lowest value (1) and Z the highest (20).

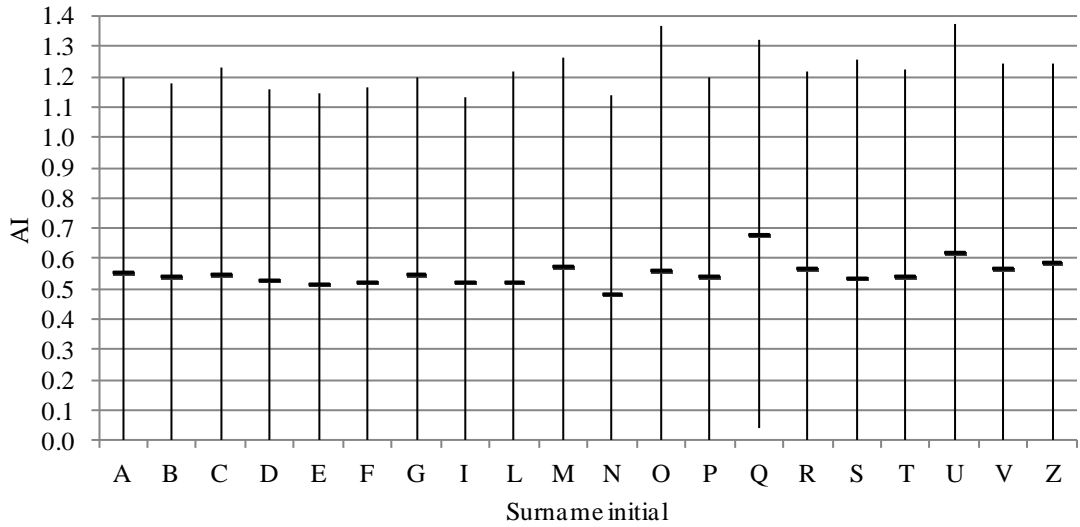

*Figure 1: $\overline{AII}$ distributions per surname initial of professors in the dataset: the horizontal dash indicates the median value, the vertical line the interquartile range*

To detect any differences across scientific communities, we repeat the same analysis at the UDA level. Table 4 presents the results of the Spearman correlation analysis in the individual UDAs. For greater robustness, in each UDA, the analysis excludes the first letters for which there are less than 30 professors. We note that the absolute values of Spearman ρ never reaches a full first decimal place, indicating that at the level of the single discipline, the hypothesis of "alphabetical discrimination" can again be rejected.

*Table 4: Spearman correlation between surname initials and $\overline{AII}$ in each UDA*

| UDA | Obs. | Spearman ρ |
|---|---|---|
| 1 - Mathematics and computer science | 2,089 | 0.004 |
| 2 – Physics | 2,102 | -0.003 |
| 3 – Chemistry | 2,774 | -0.023 |
| 4 - Earth sciences | 820 | -0.022 |
| 7 - Agricultural and veterinary sciences | 1,915 | 0.014 |
| 8 - Civil engineering | 943 | -0.046 |
| 9 - Industrial and information engineering | 3,357 | -0.009 |
| Total | 14,000 | -0.012 |

One could suspect that some kind of alphabetical advantage might instead appear among scientists achieving highly in terms of number of publications. In fact these individuals would stand to benefit more from the attention of a surname with an earlier letter. We therefore repeat the above analysis, but only for the professors in the top 10% of their respective SDS by yearly average number of publications in the 2001-2004 period. The results once again reveal an absence of correlation. As shown in Table 5 the Spearman ρ is equal to 0.007 at the overall level, while for the individual UDAs it is always very low and never meaningful.

Furthermore, we have verified the significance between the differences in values of $\overline{AII}$ for two particular groups of professors: those with surnames beginning A, B or C, compared to those with surnames beginning S, T or V (the three letters, among the bottom ones, with the highest numbers of observations). The Wilcoxon rank-sum (Mann-Whitney) test rejects the hypothesis of significant differences between the distributions of these two subgroups (*p value* 0.563). Finally, we repeat the same test for 190 specific pairs of letters:[11] for most pairs (184) the difference is statistically non-significant. At the 95% confidence level, significant differences are found for only six pairs: D vs M, F vs M, M vs N, N vs Q, N vs R, N vs V (p-values respectively of 0.030, 0.012, 0.040, 0.030, 0.030). As we can see, the more recurrent initials in these pairs are M and N, which fall exactly in the middle of the Italian alphabet, confirming that there is no advantage or disadvantage for the earlier or later initials.

*Table 5: Spearman correlation between surname initials of top scientists and $\overline{AII}$ in each UDA*

| UDA | Obs | Spearman ρ |
|---|---|---|
| 1 - Mathematics and computer science | 255 | 0.057 |
| 2 - Physics | 227 | -0.021 |
| 3 - Chemistry | 307 | -0.007 |
| 4 - Earth sciences | 110 | -0.129 |
| 7 - Agricultural and veterinary sciences | 241 | 0.024 |
| 8 - Civil engineering | 131 | -0.111 |
| 9 - Industrial and information engineering | 379 | 0.098 |
| Total | 1,650 | 0.007 |

## 4. Conclusions

The scientists belonging to fields where the practice of alphabetically ordering the publication authors is still more or less occurring, have almost certainly been afflicted by misgivings that an early surname initial enhances citability. It is no accident that some have resorted to altering their surname, to move towards the front of the byline (Kadel & Walter, 2015; Efthyvoulou, 2008). This is especially natural where evaluation processes based on citations support decisions fundamental to academic research (funding) and career progress (recruitment and advancement). Any natural doubts may have been fed by the results of some empirical investigations, such as those reported in the introduction. Apart from the "line academics", the evaluators and decision-makers would also be well-advised to pay attention to such doubts, whether their evaluations are being used in competitions for academic positions and research funds, or for broadly purposed national assessment exercises.

The few prior investigations on the subject analyzed the surname effect at publication level. What we investigate instead is the effect of surname initial on the citability of individual authors. In the sciences, findings show that researchers with the alphabetically higher surnames do not result as gathering more citations. This is true without exception across scientific communities, regardless of how diffuse is the practice of alphabetically ordering the authors' list; and independently of the number of publications that the scientist produces (the higher the number of publications by the scientists, the more evident the surname effect should be).

The restriction of the analysis to Italy, while making it possible and more precise, thanks to an authors' name disambiguation algorithm developed by the authors, should not jeopardize, all others equal, the generalization of results, as Italian articles are cited from all over the world and not just by Italian scholars. Of course, results need to be interpreted accounting for the extent to which alphabetical order is practiced: even in the disciplines that we have considered, a rather significant share of the publications do not adopt alphabetical authorship.

---

[11] The permutations of pairs possible from 20 letters are: $\frac{20!}{2! \cdot (20-2)!}=190$